\documentclass[twocolumn,showpacs,nofootinbib,prd]{revtex4-2}
\usepackage{graphicx}
\usepackage{rotating}
\usepackage{color}
\usepackage{amssymb}
\newcommand{\tpo}{${}^{3\!}P_1$}
\newcommand{\spo}{${}^{1\!}P_1$}
\begin{document}
\title{Comment on \em ``Do near-threshold molecular states mix with
neighboring \boldmath$\bar QQ$ states?''}
\author{
George~Rupp$^{1}$}
\affiliation{
$^{1}$Centro de F\'{\i}sica Te\'{o}rica de Part\'{\i}culas,
Instituto Superior T\'{e}cnico, Universidade de Lisboa,
P-1049-001 Lisboa, Portugal
}

\begin{abstract}
I comment on a paper by Christoph Hanhart and Alexey Nefediev, published in
Phys.\ Rev.\ D {\bf106}, 114003 (2022).
The authors discuss the interpretation of mesons close to 
their lowest decay threshold and present a mechanism for the formation of 
molecular states. The proposed formalism is then applied to the axial-vector
mesons $D_{s1}(2536)$ and $D_{s1}(2460)$, presenting two scenarios for the
lighter meson, namely a $D^\star K$ molecule or a compact $c\bar{s}$
state. The authors argue that the latter hypothesis requires a fine-tuning
of the mixing angle between the $J^{PC}=1^{++}$ and $J^{PC}=1^{+-}$ 
$C$-parity eigenstates. 

In this Comment I show that no such fine-tuning is needed, as demonstrated
in an article published in Phys.\ Rev.\ D {\bf84}, 094020 (2011), where a 
unitarized quark model was applied to the two $C$-parity eigenstates, coupled
to several two-meson channels including $D^\star K$. The coupled-channel
dynamics naturally leads to a mixing angle very close to the required one.
Moreover, I argue that the $D_1(2420)$ and $D_1(2430)$ axial-vectors, not
considered by the authors, as well as a lattice simulation in Phys.\ Rev.\ D
{\bf90}, 034510 (2014), also not mentioned by the authors, do not lend 
support to a molecular interpretation of $D_{s1}(2460)$. I conclude with
some more general remarks about mesons coupling to $S$-wave thresholds.
\end{abstract}

\maketitle

In Ref.~\cite{HN22}, hereafter referred to as HN22, the authors study the
axial-vector (AV) charm-strange
mesons $D_{s1}(2536)$ and $D_{s1}(2460)$ in the framework of the 
Weinberg approach to determining the nature of hadronic states. In particular,
they employ this method by formulating a simple two-channel description of
these and similar mesons, in which a compact QCD-based system with an
unspecified quark content is coupled to one two-meson channel in an
$S$-wave. From their analysis, the authors conclude that there are two
possible scenarios for the mentioned charm-strange mesons in terms of their
internal structure. In the first, so-called ``strong-coupling'' scenario, the
$D_{s1}(2460)$ is claimed to emerge as a $D^\star K$ molecule, which largely
decouples from the nearby $J^{P}=1^{+}$ quark-antiquark state, whereas the
$D_{s1}(2536)$ must then be the (dominantly) $c\bar{s}$ meson. In the second,
``weak-coupling'' scenario, both mesons will be dominantly of the $c\bar{s}$
type, resulting from a mixture of the spectroscopic states \tpo\
($J^{PC}=1^{++}$) and \spo\ ($J^{PC}=1^{+-}$), as physical charm-strange 
mesons have no definite $C$-parity. The most important point of the present
Comment is to rebut the claim in HN22 that the latter scenario requires
a fine-tuning of the mixing angle between the \tpo\ and \spo\ states, based
on a unitary coupled-channel model calculation published in Ref.~\cite{CRB11},
to be referred to as CRB11. Further arguments against a molecular
interpretation of the $D_{s1}(2460)$ will be presented as well.

\begin{figure}[!t]
\centering
\includegraphics[trim = 23mm 120mm 52mm 5mm,clip,width=8.0cm]
{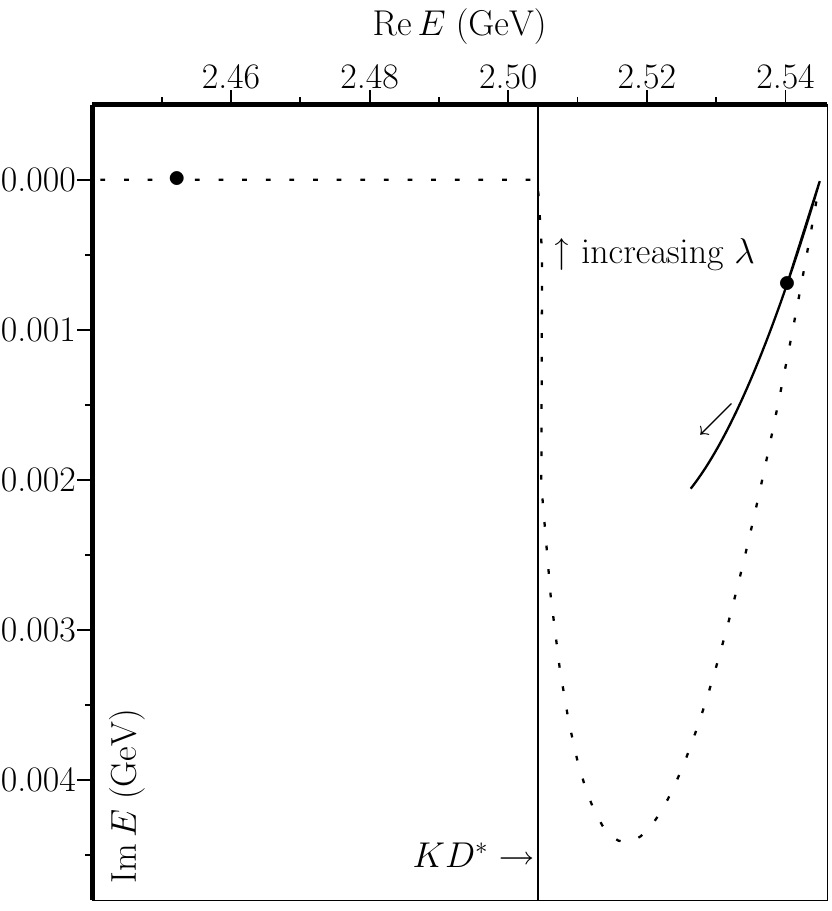}
\caption{$S$-matrix pole trajectories of the $D_{s1}(2536)$ (solid
line) and $D_{s1}(2460)$ (dotted line) as a function of the overall coupling
$\lambda$. Dots represent the physical pole positions for the same value
of $\lambda$. Figure reprinted from FIG.~3 of Ref.~\cite{CRB11}. \\[-8mm]}
\label{ds1}
\end{figure}

In CRB11, both axial-vector charm-strange ($c\bar{s}$) and charm-light
($c\bar{q}$) mesons were studied in the context of a fully unitary and
analytic multichannel model, with quark-antiquark as well as meson-meson
channels. In either case, the two spectroscopic quark-antiquark eigenstates
\tpo\ and \spo\ are coupled to all OZI-allowed vector-pseudoscalar and
vector-vector channels. Of these, only the $D^\star K$ and $D^\star\pi$
channels are kinematically open for the corresponding bare $1^{+\pm}$
$c\bar{s}$ and $c\bar{q}$ states, respectively, However, all other channels
also contribute to the real part of the physical masses through their loops.
Owing to the difference in couplings of the \tpo\ and \spo\ components to the
two-meson channels, their assumed degeneracy in the CRB11 model is lifted upon
allowing interaction with the two-meson channels. In both cases, the $S$-matrix
pole of the resulting mixture of \tpo\ and \spo\ states that is dominantly
\tpo\ shifts much more than the orthogonal combination. Focusing for the moment
on the $c\bar{s}$ system, we observe an extremely small shift of the mostly
\spo\ pole, giving rise to a $D^\star K$ decay width of the order of 1~MeV, as
well as a real part of the energy close to the mass of the $D_{s1}(2536)$
meson. On the other hand, the dominantly \tpo\ pole shifts downward by almost
100~MeV and so below the $D^\star K$ threshold, with a resulting mass close to
that of the $D_{s1}(2460)$ and zero width in the OZI-only approximation. The
two pole trajectories as a function of the overall model coupling $\lambda$ are
shown in Fig.~\ref{ds1}, reprinted from CRB11. The conclusion is that the
coupled-channel dynamics automatically leads to a mixing angle very close to
the ideal one, i.e., $\cos\Theta_{\rm AV}\approx \sqrt{2/3} \Rightarrow
\Theta_{\rm AV}\approx 35.3^\circ$. (Note that in the conventions of HN22 the
ideal mixing angle is given by $90^\circ$ minus the angle defined in CRB11.)
So absolutely no fine-tuning of this angle
is needed, as already concluded more generally in Ref.~\cite{BR04} by
diagonalizing a Hamiltonian for two degenerate bare quark-antiquark states
coupled to the continuum. Also the model of Ref.~\cite{Simonov08}, based
on a chiral quark-pion Lagrangian, supports the interpretation of the
$D_{s1}(2536)$ and $D_{s1}(2460)$ as in CRB11 and Ref.~\cite{BR04}, albeit
without generating the corresponding mixing angle dynamically.

Turning now our attention to the $c\bar{q}$ system and the PDG \cite{PDG22}
mesons $D_1(2420)$ and $D_1(2430)$, not considered in HN22, the coupled-channel
dynamics employed in CRB11 works equally well, leading to a $D_1(2420)$ state
much narrower than one would expect for a meson decaying to $D^\star\pi$ in
an $S$-wave and a lot of phase space, as well as a very broad $D_1(2430)$
resonance. Note that in the latter case the pole does not move below the ---
much lower lying --- $D^\star\pi$ threshold, which can be understood as a
consequence of an effective Adler zero in this channel. As for the broad
$D_1(2430)$ resonance, it would be very difficult to understand as a
meson-meson molecule in the philosophy of HN22, which immediately raises
the question why the $c\bar{s}$ and $c\bar{q}$ systems should be dealt with
differently in that approach. In contrast, the formalism in CRB11 allows to 
understand the complete pattern of $D_{s1}(2460)$, $D_{s1}(2536)$, $D_1(2430)$,
and $D_1(2420)$ masses and widths --- 8 observables in total ---  with only
two adjustable parameters, one of which already strongly constrained by
prior model applications. Moreover, in neither case is a fine-tuning of the
\tpo-\spo mixing angle necessary.

Let us now see what insight on AV charm mesons is provided by the lattice.
In Ref.~\cite{Lang14}, $J^P=0^+$, $1^+$, and $2^+$ charm-strange mesons
were studied by including both quark-antiquark and two-meson interpolating
fields in the simulations. In the AV case, a $c\bar{s}$ state appears below the
$D^\star K$ threshold even in the single-hadron approach, that is, without
including $D^\star K$ interpolators. However, the inclusion of $D^\star K$
scattering operators significantly improves the signal and a clear
identification of the state with the $D_{s1}(2460)$ can be made. Furthermore,
a narrow $D_{s1}(2536)$ is found in the same simulation. Therefore, 
interpreting the $D_{s1}(2460)$ as a molecular $D^\star K$ state largely
decoupled from $c\bar{s}$ is not supported by this lattice calculation. As for
the $D_1(2420)$ and $D_1(2430)$ mesons, members of the same lattice
collaboration \cite{Mohler13} carried out a similar simulation as in 
Ref.~\cite{Lang14}, obtaining results chiefly in agreement with the earlier
findings in CRB11, just like in the $c\bar{s}$ case \cite{Lang14}. Furthermore,
a recent lattice computation \cite{Lang_Wilson} by members of the Hadron
Spectrum Collaboration generally confirmed the results on the $D_1(2420)$ and
$D_1(2430)$ found in Ref.~\cite{Mohler13}. Also note the remark in a very
recent lattice talk \cite{N_Lang} about these two mesons:
\begin{quote} \em
``A stable spectrum requires both $q\bar{q}$- and meson-meson-like operators!''
\em
\end{quote} 

Some further observations are due concerning a couple of other enigmatic mesons
mentioned by the authors of HN22. Most notably there is the remarkable
$J^{PC}=1^{++}$ charmonium state $\chi_{c1}(3872)$ \cite{PDG22}, also still
called $X(3872)$, which lies practically on top of the $S$-wave
$\bar{D}^{\star0}D^0$ threshold and so has been frequently considered a
molecular candidate. In Ref.~\cite{Lang15}, members of the same lattice 
collaboration as in Refs.~\cite{Lang14,Mohler13} employed $c\bar{c}$,
$\bar{D}^{\star0}D^0$, and even tetraquark interpolators to describe the
$X(3872)$. Their definitely most important conclusion was that the
state does not survive if no $c\bar{c}$ interpolators are included, with
$\bar{D}^{\star0}D^0$ also being important in order to obtain a clear
signal close to the experimentally observed meson. The additionally included
tetraquark interpolators turn out to be practically immaterial in the
simulations. These lattice results lend support to earlier momentum-space
\cite{CRB11-2} and coordinate-space \cite{CRB13} (Ref.~[13] in HN22) 
coupled-channel model calculations. Also note that the multichannel calculation
in Ref.~\cite{CRB15}, including several $S$-wave and $D$-wave two-meson
channels, reveals an $X(3872)$ wave function dominated by a pronounced
$c\bar{c}$ core, despite an overall $\bar{D}^{\star0}D^0$ probability of about
65\% due to the extremely long tail of this wave-function component owing to
the tiny binding energy.

Another famous meson is the $D_{s0}^\star(2317)$ charm-strange scalar, which 
has also been often considered a molecule or a tetraquark state.
In Ref.~\cite{Lang13}, members of the lattice collaboration of
Refs.~\cite{Lang14,Lang15,Mohler13} described this scalar meson on a combined
basis of $c\bar{s}$ and $DK$ operators, in much the same way as the
$D_{s1}(2460)$ with $c\bar{s}$ and $D^\star K$ operators in Ref.~\cite{Lang14},
finding a state below the $DK$ threshold compatible with the physical
$D_{s0}^\star(2317)$. The inclusion of both types of interpolating fields is
crucial to obtain energy levels with small statistical uncertainties. The
results of this simulation support the earlier coupled-channel modeling of this
meson in Ref.~\cite{BR03}. In another and more recent
lattice simulation \cite{Wagner20}, including tetraquark interpolating fields
besides $c\bar{s}$ and $DK$, the $D_{s0}^\star(2317)$ emerges as a state below
the $DK$ threshold that is mostly of a quark-antiquark type, with a small $DK$
component. The tetraquark interpolators turn out to be essentially irrelevant.

Now, one should realize the special features of non-exotic $S$-wave
meson-meson scattering, in which $q\bar{q}$ states couple very strongly 
to two-meson channels. Most significantly, the light scalar
mesons have been shown to appear naturally as extra and dynamically generated
resonances in unitary coupled-channel models formulated in coordinate space
\cite{Beveren86} and momentum space \cite{BBKR06}, besides the regular 
quark-model scalars in the ballpark of 1.3--1.5~GeV. In that respect, let us
quote from HN22:
\begin{quote} \em
``Some bare poles appear below and some above the $\varphi\bar{\varphi}$
threshold. In the regime of small coupling, the poles lying above the
threshold get shifted to the complex plane and then, as the coupling
increases, their trajectories bend and reapproach the real axis. Such a
behaviour of the poles was previously discussed in the literature --- see,
for example, Refs.~[25,42,43].''
\end{quote}
Now, the authors of HN22 mentioning their Ref.~[42] (Ref.~\cite{BBKR06}
in the present Comment) in the quoted remark are mistaken, because for small
coupling the poles of the $f_0(500)$ and $K_0^\star(700)$ lie very deep in the
complex energy plane, while those of the $f_0(980)$ and $a_0(980)$ can also
not be traced back to the real axis in that limit \cite{BBKR06}. This
phenomenon of generating meson resonances via nonperturbative coupled-channel
dynamics in $S$-wave scattering \cite{Oller97} is not exclusive to the light
scalars, but can also be seen in the case of, for example, the
$D_{s0}^\star(2317)$ \cite{BR03} and the $X(3872)$ \cite{CRB11-2,CRB13}. 
To make life even more complicated, the dynamically generated $X(3872)$ pole
was shown in Ref.~\cite{CRB13} to interchange its identity with an intrinsic
pole for small changes in a model parameter, while hardly influencing the
resulting pole position. Therefore, identifying an $S$-wave meson resonance as 
dynamically generated or intrinsic can sometimes be very cumbersome, which
adds to the many complexities of meson spectroscopy.

In conclusion, I point out that all the mesons mentioned in the foregoing
share the property of coupling in an $S$-wave to their lowest or dominant
meson-meson decay threshold. What does distinguish them, though, is the precise
locations of those thresholds, which depend on the quantum numbers and
constituent quark masses in the decay products. So my assessment is that the
proximity of their masses to their lowest (or dominant) $S$-wave two-meson
decay thresholds, with the $X(3872)$ being an extreme case, is to some extent
accidental yet not entirely, in view of the undeniable role that such $S$-wave
thresholds play in locking \cite{Bugg08} or even generating \cite{Beveren86}
$S$-matrix poles.

Nevertheless, more definite conclusions can only be drawn
from precise experimental data, as the authors of HN22 suggest in the case of
the $D_{s1}(2460)$. However, the problem with their proposal to measure the
theoretically estimated width of the isospin-violating decay
$D_{s1}(2460)^+\to D_s^{\star+}\pi^0$, of the order of 100~keV in the molecular
scenario, is not only a challenge to experiment, as the authors themselves
admit. Because also the theoretical estimates, both in the molecular and
$c\bar{s}$ scenarios, are based on several assumptions, not to speak of the
inevitable $D^\star K$ loop contributions in the (dominantly) $c\bar{s}$
case as well. So I suggest electromagnetic decays like
$D_{s1}(2460)^+\to D_s^{\star+}\gamma$ and
$D_{s1}(2460)^+\to D_{s0}^\star(2317)^+\gamma$ 
as better candidates \cite{PDG22} to probe the structure of $D_{s1}(2460)$,
just like for the $X(3872)$ \cite{Simonov12,CRB15}.


\begin{thebibliography}{99} 
\bibitem{HN22}
C.~Hanhart and A.~Nefediev,
Phys.\ Rev.\ D {\bf106}, 114003 (2022)
[arXiv:2209.10165 [hep-ph]].

\bibitem{CRB11}
S.~Coito, G.~Rupp, and E.~van Beveren,
Phys.\ Rev.\ D {\bf84}, 094020 (2011)
[arXiv:1106.2760 [hep-ph]].

\bibitem{BR04}
E.~van Beveren and G.~Rupp,
Eur.\ Phys.\ J.\ C {\bf32}, 493 (2004)
[arXiv:hep-ph/0306051 [hep-ph]].

\bibitem{Simonov08}
A.~M.~Badalian, Y.~A.~Simonov, and M.~A.~Trusov,
Phys.\ Rev.\ D {\bf77}, 074017 (2008)
[arXiv:0712.3943 [hep-ph]].

\bibitem{PDG22}
R.~L.~Workman {\it et al.} [Particle Data Group],
PTEP {\bf2022}, 083C01 (2022).

\bibitem{Lang14}
C.~B.~Lang, L.~Leskovec, D.~Mohler, S.~Prelovsek, and R.~M.~Woloshyn,
Phys.\ Rev.\ D {\bf90}, 034510 (2014) 
[arXiv:1403.8103 [hep-lat]].

\bibitem{Mohler13}
D.~Mohler, S.~Prelovsek, and R.~M.~Woloshyn,
Phys.\ Rev.\ D {\bf87}, 034501 (2013) 
[arXiv:1208.4059 [hep-lat]].

\bibitem{Lang_Wilson}
Nicolas Lang and David J.~Wilson [Hadron Spectrum Collaboration],
Phys.\ Rev.\ Lett.\ {\bf129} (2022) 252001
[arXiv:2205.05026 [hep-ph]].

\bibitem{N_Lang}
Nicolas Lang,
{\it ``Charmed meson resonances from Lattice QCD''},
Talk at Workshop Excited QCD 2024, Benasque, Spain, 14--20 January 2024;
see page 20: \\
{\tt\scriptsize
https://www.benasque.org/2024eqcd/talks\_contr/185\_5-talk.pdf} \\[-3mm]

\bibitem{Lang15}
M.~Padmanath, C.~B.~Lang, and S.~Prelovsek,
Phys.\ Rev.\ D {\bf92}, 034501 (2015) 
[arXiv:1503.03257 [hep-lat]].

\bibitem{CRB11-2}
S.~Coito, G.~Rupp, and E.~van Beveren,
Eur.\ Phys.\ J.\ C {\bf71}, 1762 (2011)
[arXiv:1008.5100 [hep-ph]].

\bibitem{CRB13}
S.~Coito, G.~Rupp, and E.~van Beveren,
Eur.\ Phys.\ J.\ C {\bf73}, 2351 (2013)
[arXiv:1212.0648 [hep-ph]].

\bibitem{CRB15}
M.~Cardoso, G.~Rupp, and E.~van Beveren,
Eur.\ Phys.\ J.\ C {\bf75}, 26 (2015)
[arXiv:1411.1654 [hep-ph]].

\bibitem{Lang13}
D.~Mohler, C.~B.~Lang, L.~Leskovec, S.~Prelovsek, and R.~M.~Woloshyn,
Phys.\ Rev.\ Lett.\ {\bf111}, 222001 (2013)
[arXiv:1308.3175 [hep-lat]].

\bibitem{Wagner20}
C.~Alexandrou, J.~Berlin, J.~Finkenrath, T.~Leontiou, and M.~Wagner,
Phys.\ Rev.\ D {\bf101}, 034502 (2020)
[arXiv:1911.08435 [hep-lat]].

\bibitem{BR03}
E.~van Beveren and G.~Rupp,
Phys.\ Rev.\ Lett.\ {\bf91}, 012003 (2003)
[arXiv:hep-ph/0305035].

\bibitem{Beveren86}
E.~van Beveren, T.~A.~Rijken, K.~Metzger, C.~Dullemond, G.~Rupp,
and J.~E.~Ribeiro,
Z.\ Phys.\ C {\bf30}, 615 (1986)
[arXiv:0710.4067 [hep-ph]].

\bibitem{BBKR06}
E.~van Beveren, D.~V.~Bugg, F.~Kleefeld, and G.~Rupp,
Phys.\ Lett.\ B {\bf641}, 265 (2006)
[arXiv:hep-ph/0606022 [hep-ph]].

\bibitem{Oller97}
J.~A.~Oller and E.~Oset,
Nucl.\ Phys.\ A {\bf620}, 438 (1997)
[Erratum-ibid {\bf652}, 407 (1999)]
[arXiv:hep-ph/9702314 [hep-ph]].

\bibitem{Bugg08}
D.~V.~Bugg,
J.\ Phys.\ G {\bf35}, 075005 (2008)
[arXiv:0802.0934 [hep-ph]].

\bibitem{Simonov12}
A.~M.~Badalian, V.~D.~Orlovsky, Y.~A.~Simonov, and B.~L.~G.~Bakker,
Phys.\ Rev.\ D {\bf85}, 114002 (2012)
[arXiv:1202.4882 [hep-ph]].

\end{thebibliography}
\end{document}